\documentclass[sigconf]{acmart}

\AtBeginDocument{%
  \providecommand\BibTeX{{%
    \normalfont B\kern-0.5em{\scshape i\kern-0.25em b}\kern-0.8em\TeX}}}

\copyrightyear{2020} 
\acmYear{2020} 
\setcopyright{rightsretained} 
\acmConference[ICSE '20 Companion]{42nd International Conference on Software Engineering Companion}{May 23--29, 2020}{Seoul, Republic of Korea}
\acmBooktitle{42nd International Conference on Software Engineering Companion (ICSE '20 Companion), May 23--29, 2020, Seoul, Republic of Korea}\acmDOI{10.1145/3377812.3382142}
\acmISBN{978-1-4503-7122-3/20/05}

\begin{document}

\title{DCO Analyzer: Local Controllability and Observability Analysis and Enforcement of Distributed Test Scenarios}

\author{Bruno Lima}
\orcid{0000-0003-2572-047X}
\affiliation{%
  \institution{INESC TEC and Faculty of Engineering, University of Porto}
  \streetaddress{Rua Dr. Roberto Frias, s/n}
  \city{Porto}  
  \postcode{4200-465}
  \country{Portugal}
}
\email{bruno.lima@fe.up.pt}

\author{Jo{\~{a}}o Pascoal Faria}
\affiliation{%
  \institution{INESC TEC and Faculty of Engineering, University of Porto}
  \streetaddress{Rua Dr. Roberto Frias, s/n}
  \city{Porto}  
  \postcode{4200-465}
  \country{Portugal}
}
\email{jpf@fe.up.pt}


\begin{abstract}
To ensure interoperability and the correct behavior of heterogeneous distributed systems in key scenarios,
it is important to conduct automated integration tests, based on distributed test components (called local testers) that are deployed close to the system components to simulate inputs from the environment and monitor the interactions with the environment and other system components. We say that a distributed test scenario is locally controllable and locally observable if test inputs can be decided locally and conformance errors can be detected locally by the local testers, without the need for exchanging coordination messages between the test components during test execution (which may reduce the responsiveness and fault detection capability of the test harness). DCO Analyzer is the first tool that checks if distributed test scenarios specified by means of UML sequence diagrams exhibit those properties, and automatically determines a minimum number of coordination messages to enforce them. 

The demo video for DCO Analyzer can be found at \url{https://youtu.be/LVIusK36_bs}.

\end{abstract}




\maketitle

\section{Introduction}
To ensure interoperability and the correct end-to-end behavior of heterogeneous distributed systems in a growing number of domains, such as IoT for e-health and smart cities, it is important to conduct integration tests that verify the interactions with the environment and between the system components in key scenarios. 

Although there are several tools for distributed systems testing, an analysis of the state of practice \cite{lima2016survey} identified the need for new approaches and tool sets to meet the needs of the industry, namely for the integration testing of heterogeneous distributed systems. 

The automation of such integration tests is challenging, because it requires that test components are also distributed, with local testers deployed close to the system components, coordinated by a central tester. In such a test architecture, it is important to minimize the communication overhead between the test components, to maximize the responsiveness and fault detection capability of the test harness. We say that a distributed test scenario is locally observable and locally controllable if conformance errors can be detected locally and test inputs can be decided locally, respectively, by the local testers, without the need for exchanging coordination messages between the test components during test execution. In fact a recent study \cite{garousi2018survey} shows that the two most often mentioned factors affecting testability are observability and controlability. 

In this paper, we introduce DCO Analyzer, a tool for analyzing local observability and controllability properties of distributed test scenarios specified by means of UML sequence diagrams (SDs), and automatically recommend a minimum set of coordination messages that can guarantee those properties. DCO Analyzer can be found at \url{https://brunolima.info/DCOANALYZER/}

The tool can be useful not only for testers, but also for system designers, because local observability and controllability problems are often due to system design flaws or incomplete specifications. 

\section{Background}

\begin{figure}[h]
  \centering
  \includegraphics[width=\linewidth]{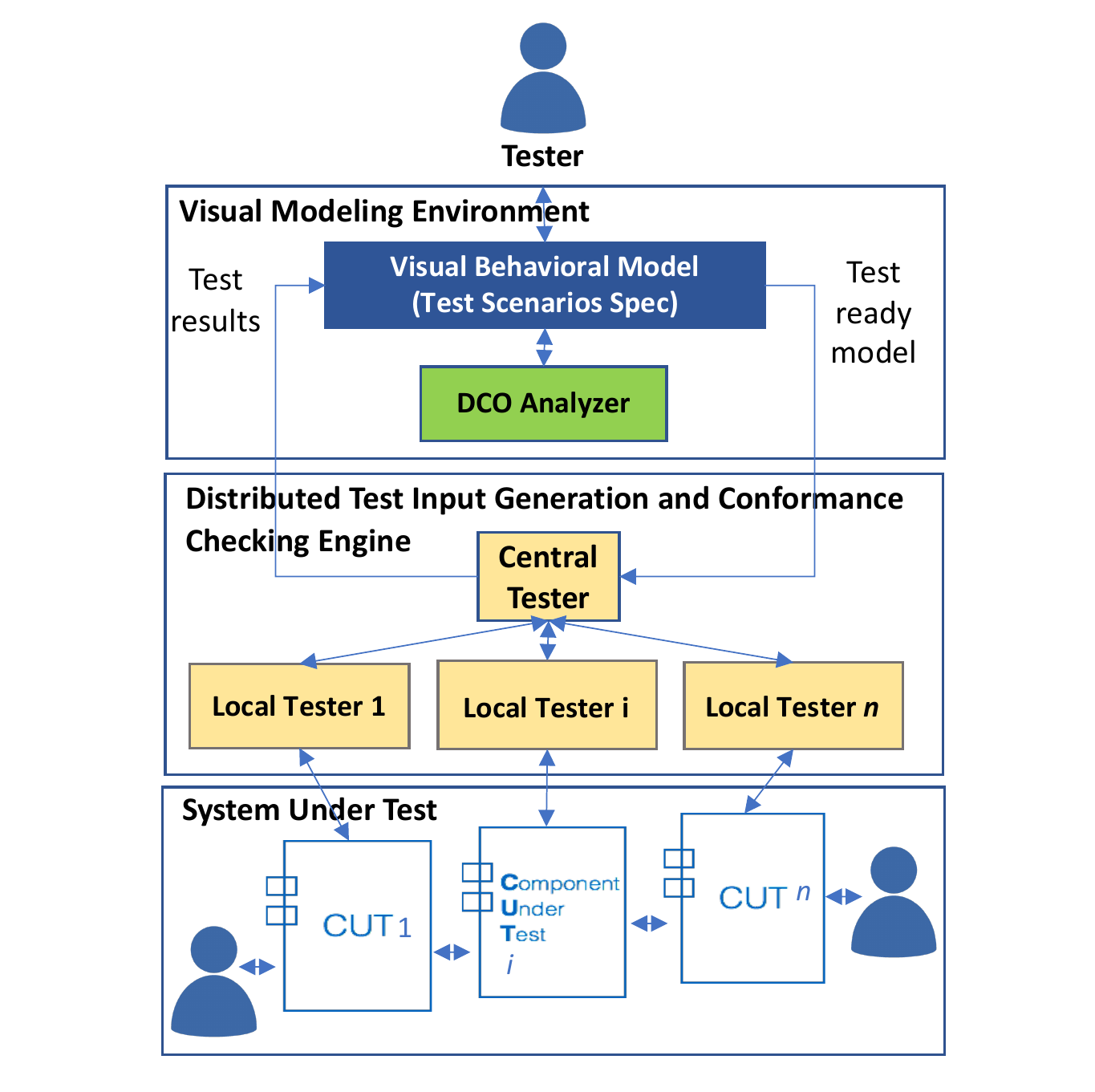}
  \caption{Model-based integration testing architecture}
  \label{fig:arch}
\end{figure}

In previous work  \cite{Lima2016}, we have proposed an online model-based testing (MBT) approach and architecture for automating the integration testing of end-to-end services in distributed and heterogeneous systems, comprising a visual modeling environment and a distributed test input generation and conformance checking engine, as depicted in Figure \ref{fig:arch}. In that approach, the only manual activity required from the tester is the description of the participants and behavior of the services under test (SUT) with UML SDs (together with mapping information between the model and the implementation), which are automatically translated to a representation amenable for efficient test input generation and conformance checking at runtime. In the case of parametrized scenarios, test data to instantiate the scenario parameters may be provided by the user or generated automatically.

The distributed test input generation and conformance checking engine comprises a set of local testers, deployed close to the system components under test (CUT), coordinated by a central tester. Local testers deployed close to CUT that interact with the environment may act both as \textit{test drivers} (simulating inputs from the environment) and \textit{test monitors} (monitoring outputs to the environment and messages exchanged with other CUT). Local testers deployed close to CUT that do not interact with the environment may act just as test monitors. 
The central and local testers are SUT independent, but local testers may need a platform adapter. 

If the given test scenario is locally observable and locally controllable (or was refined to meet those properties), after the central tester sets up and initiates the local testers, no communication between the test components is needed during test execution (besides the minimal set of coordination messages possibly added in the refinement step), and the central tester only needs to receive a verdict from each local tester at the end of successful test execution or as soon as an error is detected.
Based on the information received initially from the central tester (which need to include a specification of the valid local traces), local testers acting as test drivers should be able to decide when and what test inputs to inject (\textit{distributed test input generation}); local testers acting as test monitors should be able to check the conformance of observed events with respect to the expectations (\textit{distributed conformance checking}).

The visual modeling environment includes an "observability and controllability analysis and enforcement" component (here implemented by DCO Analyzer), for checking and refining the test scenarios specified by the tester, prior to test execution.

To our knowledge, although observability and controllability have been addressed by other authors \cite{Hierons2012, Mitchell, hierons2012using, Boroday} in the context of distributed systems testing, previous work did not address the automatic generation of coordination messages to enforce the properties of local observability and controllability in the context of integration testing.

\section{DCO Analyzer}

\subsection{Architecture and functionalities}

\begin{figure}[h]
  \centering
  \includegraphics[width=\linewidth]{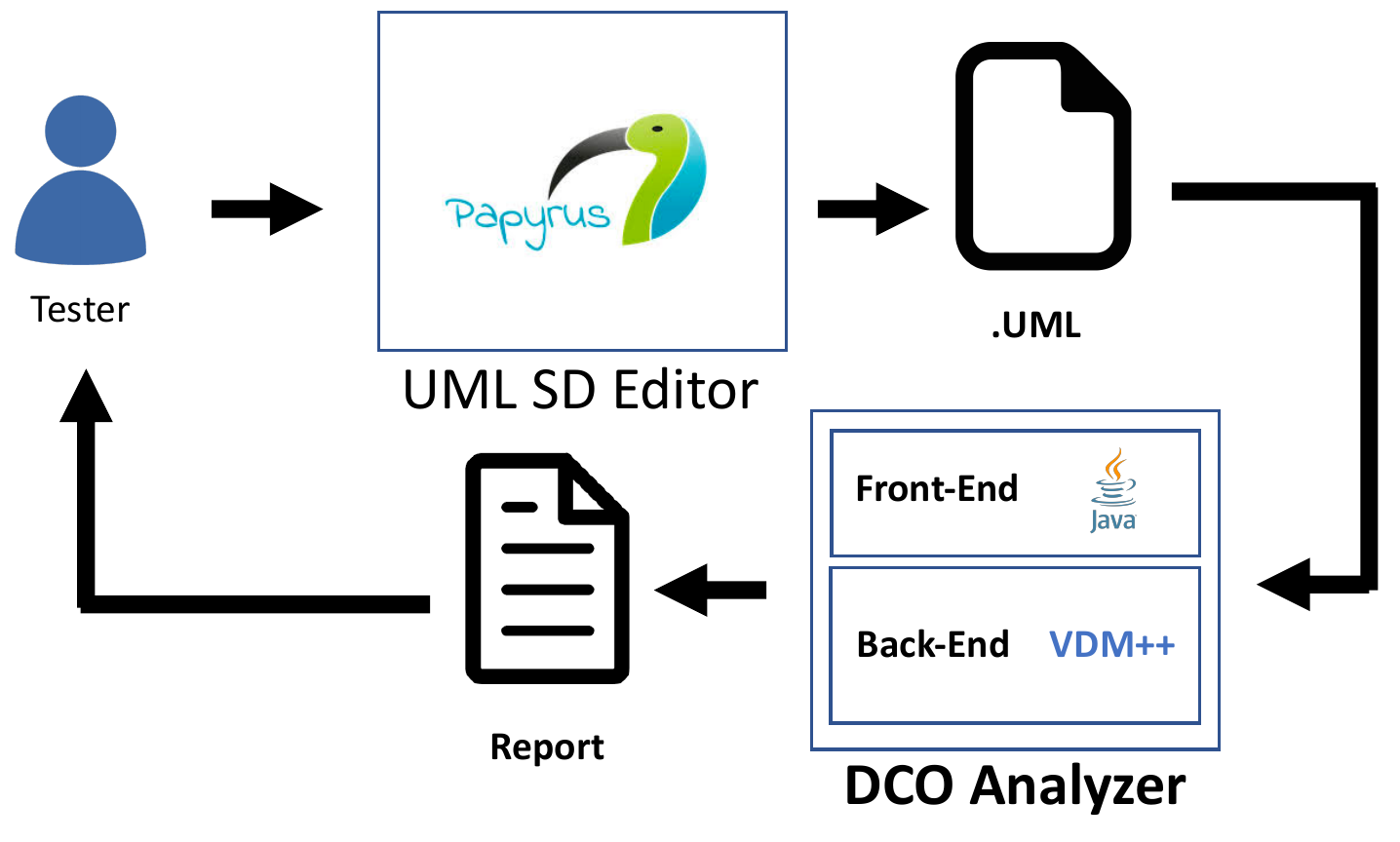}
  \caption{DCO Analyzer Overview}
  \label{fig:overview}
\end{figure}

DCO Analyzer is an application developed in Java \cite{Java} and VDM++\cite{durr1992vdm} to analyze UML SDs that represent distributed system test scenarios. As depicted in Figure \ref{fig:overview}, the user can use any visual editor of UML SDs (e.g. PAPYRUS\footnote{https://papyrusuml.wordpress.com}) and then upload the created diagrams to DCO Analyzer.

Internally, DCO Analyzer is composed of a front-end, developed in Java, and a back-end, developed in VDM++. The front-end is responsible for receiving and parsing .uml files describing UML SDs, verifying their conformance with the UML metamodel \cite{OMG2017}, and converting them into the formal representation expected by the back-end (VDM++ data structures).

The back-end analyzes the following properties:

\begin{itemize}
	\item \textbf{Valid Traces:} Set of valid global traces defined by the SD; 
	\item \textbf{Unintended Traces:} Set of invalid global traces caused by locally valid decisions (representing violations of local controllability);
	\item \textbf{Locally Uncheckable Traces:} Set of invalid global traces that cannot be verified locally (representing violations of local observability); 
	\item \textbf{Local Controllability:} The diagram is locally controllable if there are no unintended traces;
	\item \textbf{Local Observability:} The diagram is locally observable if there are no locally uncheckable traces;
	\item \textbf{Coordination Messages:} If the SD is not locally controllable and/or not locally observable, a minimum set of coordinating messages will be generated to make the diagram locally observable and controllable, whilst preserving the valid local traces at each lifeline (except for the added coordinations events).
\end{itemize}

The first 4 properties are determined following the procedures described in \cite{Amost2017}. The algorithm used to discover minimum sets of coordination messages is described in \cite{lima2019local}. The algorithm uses the locations of local observability and local controllability problems (locations where the locally uncheckable or unintended traces deviate from valid traces) as hints for points where coordination messages need to be inserted.

\subsection{Usage, examples and implications}

\begin{figure}[h]
  \centering
  \includegraphics[width=\linewidth]{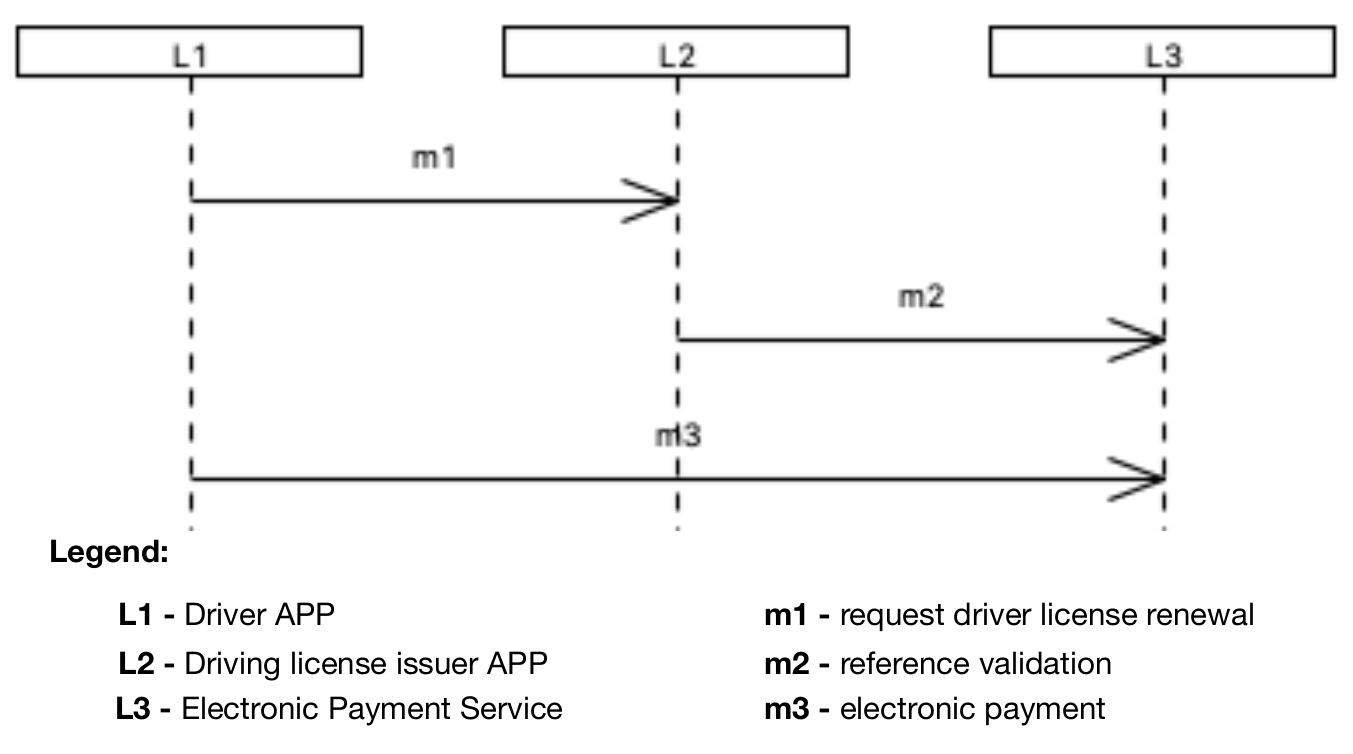}
  \caption{Initial SD in Papyrus}
  \label{fig:BasicDiagram}
\end{figure}

Figure \ref{fig:BasicDiagram} shows the SD of an online driving license renewal system (greatly simplified for illustration purposes) drawn with the PAPYRUS tool. 

\begin{figure}[h]
  \centering
  \includegraphics[width=\linewidth]{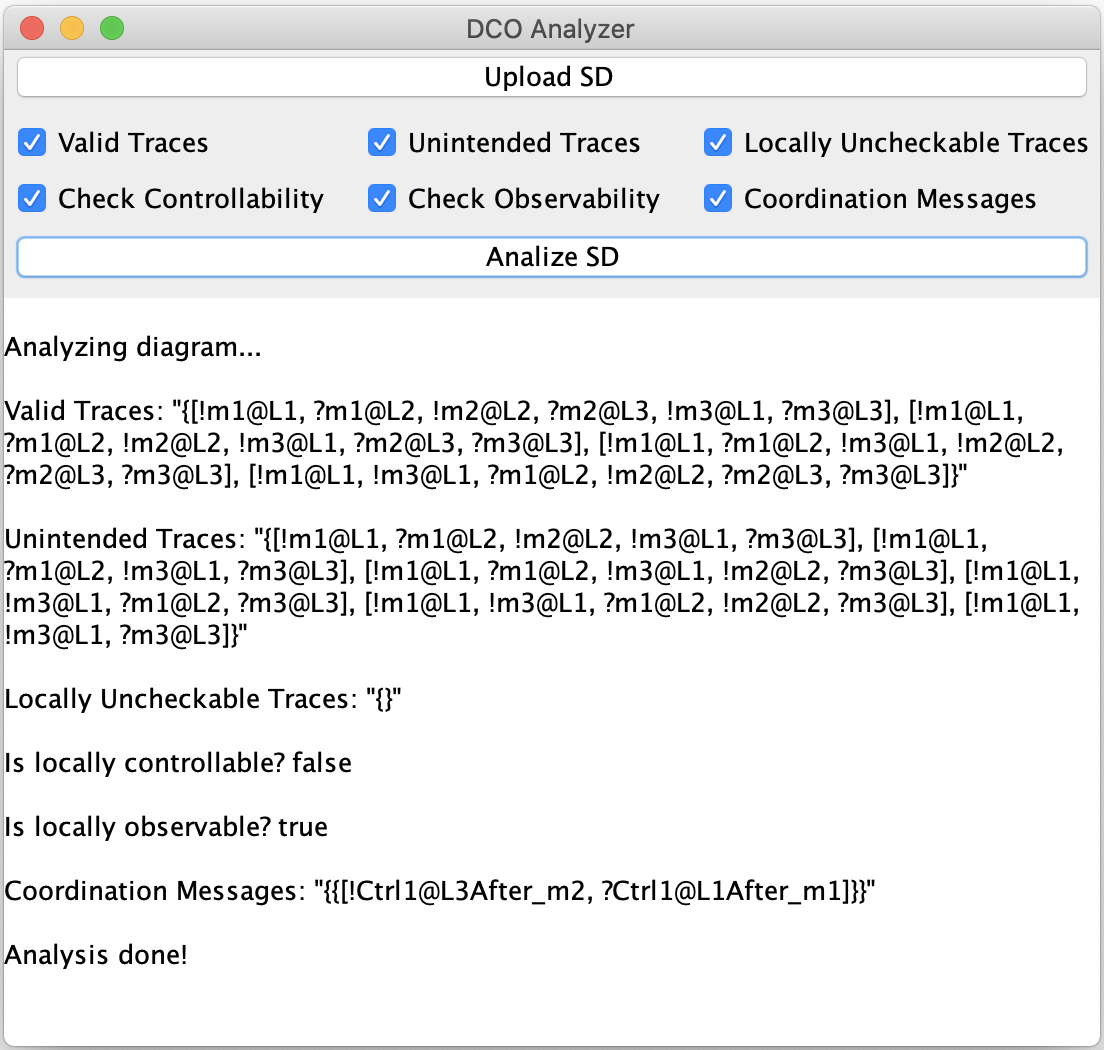}
  \caption{DCO Analyzer output}
  \label{fig:ToolOutput}
\end{figure}

From this diagram our tool is then able to analyze the properties previously described. Figure \ref{fig:ToolOutput} shows the output of the tool if requested to analyze all properties.

In the output, a set of traces is represented between $\{...\}$, a trace (sequence of events) is represented between $[...]$,  the emission of a message $m$ by a lifeline $L$ is represented as $!m@L$, and the reception of a message $m$ at a lifeline $L$ is represented as $?m@L$. 

In this example, our tool was able to detect that the given diagram is not locally controllable, indicating six unintended traces. These unintended traces are related to the possibility of the electronic payment message ($m3$) being received  by the electronic payment service ($L3$) before the reference validation message ($m2$).

In order to help the user to make this diagram locally controllable, our tool suggests adding a coordination message ($Ctrl1$) between the Electronic Payment Service ($L3$) and the Driver APP ($L2$), after messages $m2$ (at $L3$) and $m1$ (at $L1$).
In practice, such message might represent a payment authorization confirmation message, thereby ensuring that payment can only be made after the payment reference has been validated.

\begin{figure}[h]
  \centering
  \includegraphics[width=\linewidth]{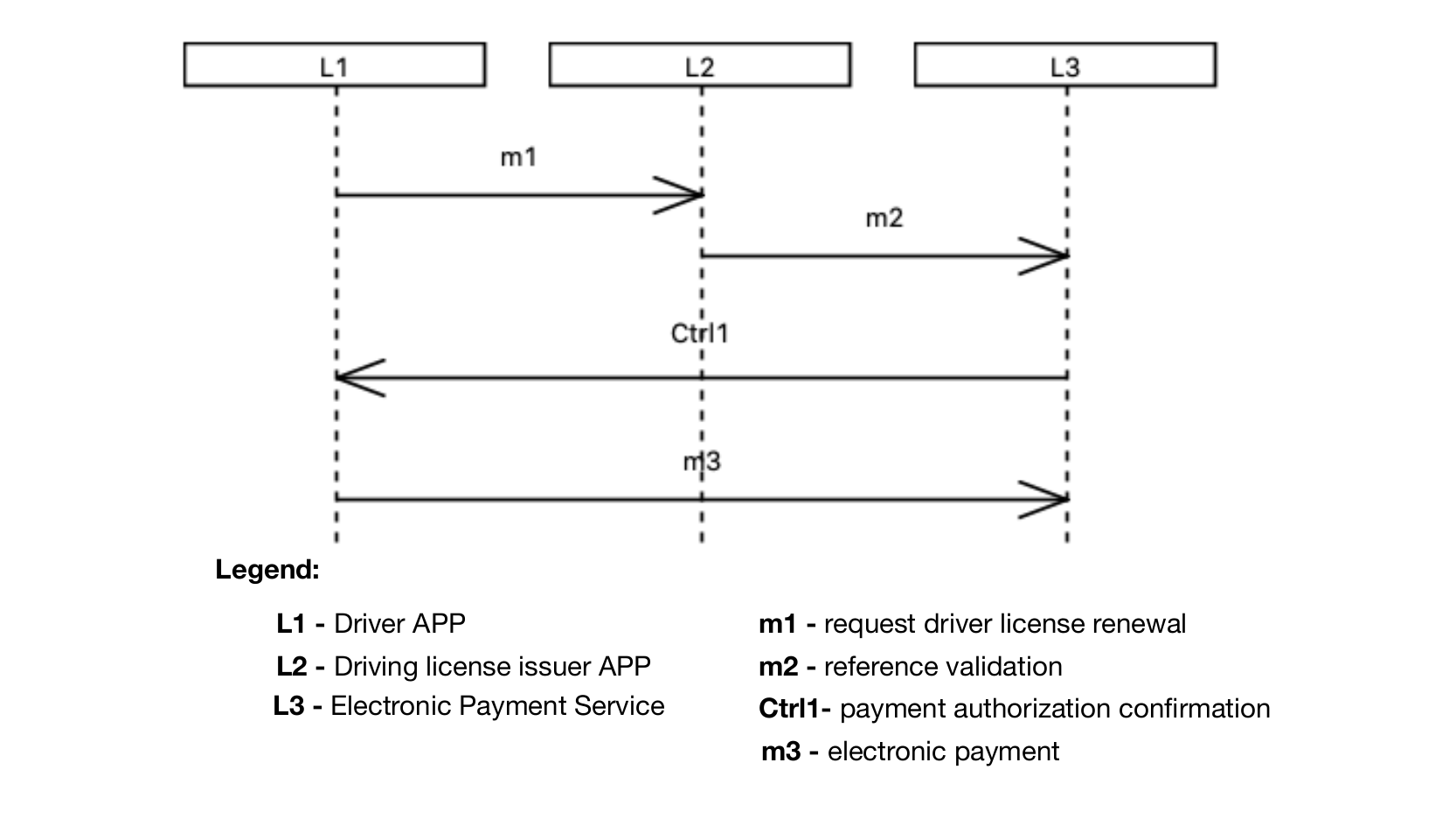}
  \caption{Refined SD in Papyrus}
  \label{fig:BasicDiagramCorrected}
\end{figure}

With this suggestion, the user can then refine the SD as shown in Figure \ref{fig:BasicDiagramCorrected}.

The suggestion given by our tool can be used in several ways:
\begin{itemize}
\item{the suggested message is actually implemented in the SUT, so the SD is just modified to include it (\textit{incomplete specification}); }

\item{the SUT is redesigned to incorporate the suggested message, and the SD is updated accordingly (\textit{design flaw}); }

\item{the system design is not changed, so the suggested message is marked as a \textit{test coordination message} to be exchanged between the test components during test execution (e.g., between a test monitor co-located with $L3$ and a test driver co-located with $L1$).}
\end{itemize}

The coordination messages suggested by DCO Analyzer are also useful in a centralized testing approach, if interpreted as general ordering constraints. In fact, in a centralized testing approach, all the events observed by the local testers are communicated to the central tester, that checks their conformance with the specification and decides the next test inputs to be injected by the local testers. In the current example, assuming that $m3$ is to be injected by the test driver at $L1$,  the central tester has to decide when $m3$ can be safely injected. The coordination message suggested by DCO Analyzer, interpreted as a general ordering constraint between $?m2@L3$ and $!m3@L1$, precisely answers that question: $m3$ can be safely injected at $L1$ after $m2$ is received at $L3$.

\begin{figure}[h]
  \centering
  \includegraphics[width=\linewidth]{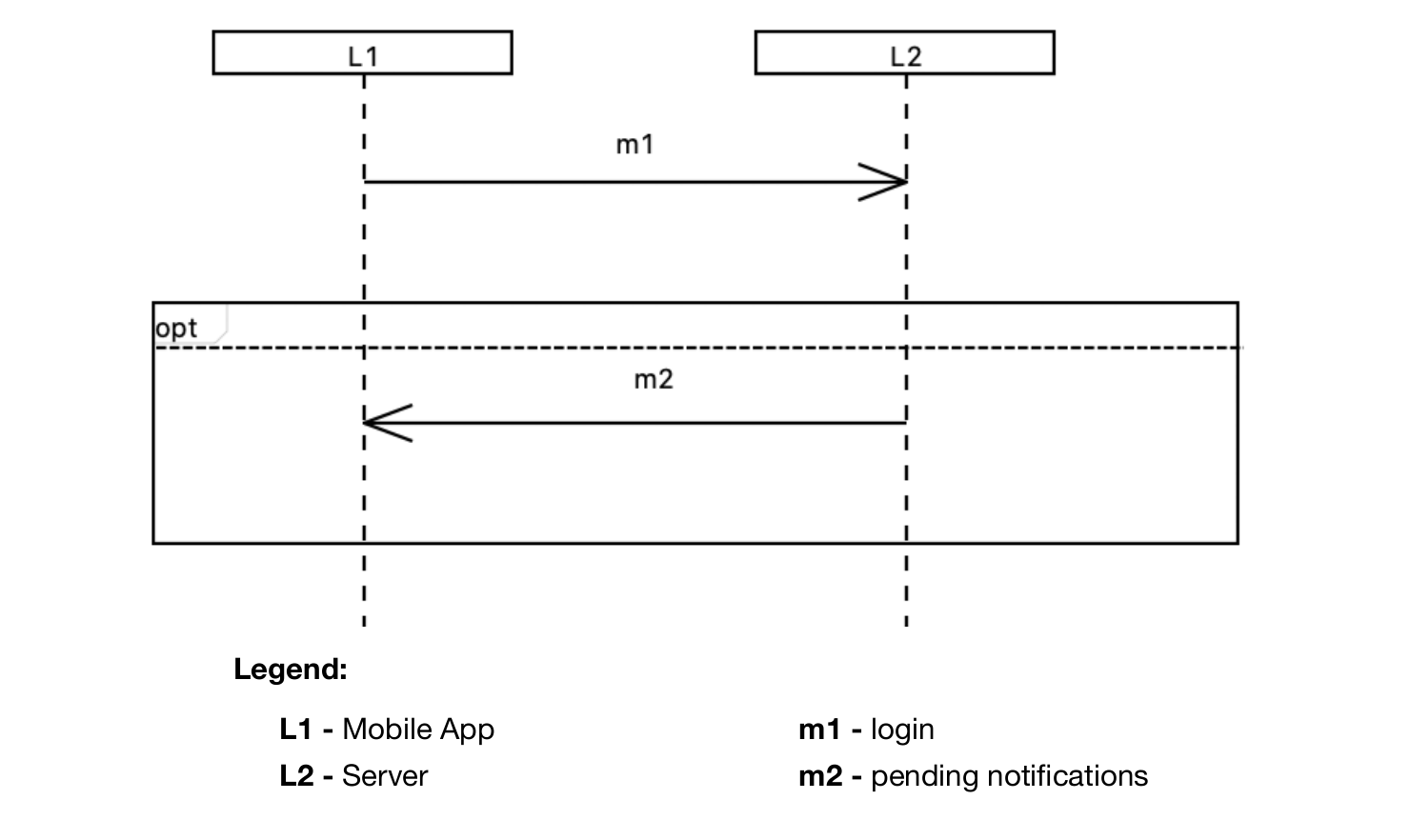}
  \caption{Example of a not locally observable scenario}
  \label{fig:notificationexample}
\end{figure}

Another example, illustrating a local observability problem, is shown in Figure \ref{fig:notificationexample}.

This diagram represents the login scenario of a mobile application, where the user, after login, can receive pending notifications since the last time the application was connected to the server. By analyzing the diagram with our tool it is possible to detect that local testers are unable to locally detect the execution trace $[!m1@L1, ?m1@L2, !m2@L2]$, which corresponds to the case where the message $m2$ is sent but lost. Such loss will not be detected as an error at $L1$ because not receiving $m2$ is also a valid behavior at $L1$. The solution to this problem recommended by the DCO Analyzer is to place a coordination message between $L1$ and $L2$ upon receipt of $m2$ in $L1$. Such message can be interpreted as an acknowledgment message; if $m2$ is lost (or the acknowledgment message is lost), then a problem will be detected at $L2$. 

More complex SDs are also supported, namely SDs with other control flow variants (\texttt{alt} and \texttt{loop} combined fragments).

\subsection{Limitations}

DCO Analyzer is fully able to find local controllability and observability problems in time constrained SDs, but its ability to enforce those properties in time constrained SDs is limited. To solve part of those limitations, we intend to investigate complementary techniques to fix local controllability and observability problems based on the addition of coordination time constraints, instead of coordination messages.

\section{Validation}

In order to validate our tool, tests were performed using the examples described in \cite{lima2019local}. These examples were chosen because they cover different features of UML, including various combined fragments and interaction operands. The results showed that our tool was able to detect controllability and observability problems and correctly generate coordination messages to fix those problems.

In addition to this initial validation, we intend to conduct an assessment with real case studies from industry partners to confirm that our tool is capable of detecting and fixing all observability and controllability issues. We also intend to understand if the outputs provided by DCO Analyzer are easily understood. This feedback from industrial partners will be very useful to suggest improvement for future versions of our tool.

\section{Conclusions}
We have presented DOC Analyzer, a controllability and observability analysis tool for distributed systems testing that is able to analyze distributed test scenarios specified by means of UML sequence diagrams, determine violations to those properties, and determines a minimum number of coordination messages to enforce them. As controllability and observability are important properties that directly affect testability (both in distributed and centralized testing approaches), we believe this tool will help testers and system designers in their daily tasks in order to define better test scenarios and make them ready for efficient test execution.

DCO Analyser was validated using a set of examples that covered the most used control flow variants in UML SDs. Future validation using test scenarios provided by industry partners is planned, not only to understand if the feedback provided by the tool is easily understood by users, but also to try to observe feedback that helps define features that may be added in future versions of the tool. We also plan to investigate controllability enforcement algorithms based on the addition of coordination time constraints (to coordinate the instants of intervention of different lifelines), instead of coordination messages.

\begin{acks}
This work was financed by the Portuguese Foundation for Science and Technology (FCT), under research grant SFRH/BD/115358/2016. 
\end{acks}

\bibliographystyle{ACM-Reference-Format}
\bibliography{sample-bibliography}


\end{document}